\let\mypdfximage\pdfximage
\def\pdfximage{\immediate\mypdfximage}

\documentclass[journal,twocolumn,9pt]{IEEEtran1}

\makeatletter
\def\markboth#1#2{\def\leftmark{\@IEEEcompsoconly{\sffamily}\MakeUppercase{\protect#1}}%
\def\rightmark{\@IEEEcompsoconly{\sffamily}\MakeUppercase{\protect#2}}}
\makeatother

\let\labelindent\relax

\usepackage{times}
\usepackage{marvosym}
\usepackage{longtable}
\usepackage{graphics}
\usepackage{graphicx}
\usepackage{caption}
\usepackage[english]{babel}
\usepackage{subfig}
\usepackage{epsfig}
\usepackage{color}
\usepackage{multirow}
\usepackage{mathrsfs}
\usepackage{textcomp}
\usepackage{amsfonts}
\usepackage{amsmath}
\usepackage{amssymb}
\usepackage{nccmath}
\usepackage[numbers,square,sort&compress,comma]{natbib}
\usepackage{chapterbib}
\usepackage[ruled,vlined,linesnumbered]{algorithm2e}
\usepackage{setspace}
\usepackage{verbatim}
\usepackage{wrapfig}
\usepackage{dblfloatfix}
\usepackage{comment}
\usepackage[strict]{changepage}
\usepackage{ragged2e}
\usepackage{microtype}
\usepackage{enumitem}
\usepackage{nccmath}
\usepackage{hyperref}
\usepackage{doi}
\usepackage{afterpage}
\usepackage{multirow,bigdelim}
\usepackage{booktabs}
\usepackage{color}
\usepackage[outline]{contour}
\usepackage{xcolor}
\usepackage{fancyhdr}
\usepackage{comment}
\usepackage{nopageno}

\setlist{parsep=0pt,listparindent=\parindent}

\DeclareCaptionFont{singlespacing}{\setstretch{1}}
\renewcommand \thesection{\arabic{section}}

\hypersetup{bookmarks=false,bookmarksopen=false,pdfpagemode=empty,pdfstartview=}

\title{\vspace{-0.55cm}\singlespacing\Large\textbf{AN INFORMATION-THEORETIC APPROACH FOR AUTOMATICALLY DETERMINING\vspace{0.01cm}\\ THE NUMBER OF STATE GROUPS WHEN AGGREGATING MARKOV CHAINS}\vspace{0.05cm}}
\author{\fontdimen2\font=2.2pt \selectfont Isaac J. Sledge${}^1$,\, \emph{Member, IEEE} \qquad Jos\'{e} C. Pr\'{i}ncipe${}^{2,3}$,\, \emph{Life Fellow, IEEE}\vspace{0.3cm}\\
${}^1$ Advanced Signal Processing and Automated Target Recognition Branch, Naval Surface Warfare Center---Panama City\\
${}^2$ Department of Electrical and Computer Engineering, University of Florida\\
${}^3$ Department of Biomedical Engineering, University of Florida\vspace{-0.2cm}
\thanks{\hspace{-0.25cm}\rule{4.5cm}{0.5pt}\newline \fontdimen2\font=1.55pt\indent This work was funded by ONR grants N00014-15-1-2013 and N00014-14-1-0542.  The first author was additionally funded by a UF Research Fellowship, a UF Robert C. Pittman Fellowship, and an ASEE NREIP Fellowship.}%
}
\begin{document}
\bstctlcite{IEEEexample:BSTcontrol}

\maketitle
\thispagestyle{empty}
\RaggedRight
\parindent=1.5em
\fontdimen2\font=2.1pt
\vspace{-0.35cm}
\IEEEpeerreviewmaketitle
\allowdisplaybreaks
\setstretch{0.9}

\vspace{-0.3cm}\section*{\textbf{ABSTRACT}}

A fundamental problem when aggregating Markov chains is the specification of the number of state groups.  Too few state groups may fail to sufficiently capture the pertinent dynamics of the original, high-order Markov chain.  Too many state groups may lead to a non-parsimonious, reduced-order Markov chain whose complexity rivals that of the original.  In this paper, we show that an augmented value-of-information-based approach to aggregating Markov chains facilitates the determination of the number of state groups.  The optimal state-group count coincides with the case where the complexity of the reduced-order chain is balanced against the mutual dependence between the original- and reduced-order chain dynamics.

{{\small{\emph{\textbf{Index Terms}}}}---Aggregation, model reduction, Markov chains, information theory}

\section*{\textbf{1.$\;\;$INTRODUCTION}}\addtocounter{section}{1}

Markov models have been widely adopted in a variety of disciplines.  Part of their appeal is that the application and simulation of such models is rather efficient, provided that the corresponding state-space has a small to moderate size.  Dealing with large state spaces is often troublesome, in comparison, as it may not be possible to adequately and efficiently simulate the underlying models \cite{ArrudaEF-conf2009a,AldhaheriRW-jour1991a,RenZ-jour2005a,SunT-jour2007a,JiaQS-jour2011a}.  

A means of rendering the simulation of large-scale models tractable is crucial for many applications.  One way to do this is by reducing the overall size of the Markov-chain state space via aggregation \cite{AokiM-jour1978a}.  Aggregation typically entails either defining and utilizing a function to partition nodes in the probability transition graph associated with the large-scale chain.  Groups of nodes, which are related by the their inter-state transition probabilities and have strong interactions, are combined and treated as a single aggregated node in a new graph.  This results in a lower-order chain with a reduced state space.  A stochastic matrix for the lower-order chain is then specified, which describes the transitions from one super-state to another.  This stochastic matrix should roughly mimic the dynamics of the original chain despite the potential loss in information \cite{DengK-conf2009a,DengK-jour2011a,GeigerBC-jour2015a}.

There are a variety of methods for aggregating Markov chains.  Some of the earliest work exploited the strong-weak interaction structure of nearly completely decomposable Markov chains to obtain reduced-order approximations \cite{AldhaheriRW-conf1989a,KotsalisG-conf2003a,SimonHA-jour1961a,CourtoisPJ-jour1975a}.  Both uncontrolled \cite{PervozvanskiiAA-jour1974a,GaitsgoriVG-jour1975a} and controlled Markov chains \cite{TeneketzisD-conf1980a,DelebecqueF-jour1981a} have been extensively studied in the literature.

In this paper, we consider an approach for aggregating nearly-completely-decomposable Markov chains, which is composed of two information-theoretic processes \cite{PrincipeJC-book2010a}.  The first process entails quantifying the dissimilarity of nodes in the original and reduced-order probability transition graphs, despite the difference in state space sizes.  The second process involves iteratively partitioning similar nodes without explicit knowledge of the number of groups.  For this second process, we consider the use of an information-theoretic criterion known as the value of information \cite{StratonovichRL-jour1965a,StratonovichRL-jour1966a} to efficiently partition the probability transition graph.  The value of information is a constrained, modified-free-energy-difference criterion that describes the maximum benefit associated with a given quantity of information in order to minimize the average distortion \cite{SledgeIJ-jour2017b,SledgeIJ-jour2017c,SledgeIJ-jour2017a}.  It is an optimal, non-linear conversion between information, usually in the Shannon sense \cite{CoverTM-book2006a}, and either costs or utilities, in the von-Neumann-Morgenstern sense \cite{vonNeumannJ-book1953a}.

Optimizing the value of information in a grouped-coordinate-descent manner yields a single free parameter that represents the effect of the information bound.  Increasing this parameter from some base value yields a hierarchy of partitions that monotonically decrease the modified-free-energy difference.  Each hierarchy element corresponds to a partition with an increasing information bound amount and hence a potentially increasing number of state groups.  Finer-scale group structure in the transition matrix is captured as the parameter value rises.  After some free-parameter value threshold, however, there are diminishing returns on the aggregation quality.  Finding this threshold, in a completely data-driven fashion, is thus crucial: over-partitioning the original chain leads to a non-parsimonious representation that can be computationally expensive to evaluate.

To automatically discern the optimal number of state groups for arbitrary Markov chains, we modify the value-of-information cost function and hence its updates.  For this modified criterion, the first update step trades off the divergence of the original and aggregated chain transition probabilities against the mutual dependence between the original and aggregated chain dynamics.  The remaining updates then trade off between how well the reduced-order model dynamics are compressed versus how much information about the original-chain dynamics is retained.  That is, it follows an information-bottleneck model \cite{TishbyN-conf1999a} with the added effect that the resulting Markov chain resembles the original.  Through update schedule, we are guaranteed that the negative modified-free-energy-difference curve is, in general, convex.  We can therefore find a maximal value on the curve.  This value corresponds to the situation where the aggregation process begins to overfit to noise in the Markov-chain transition dynamics versus the underlying nearly-completely-decomposable structure if more state groups are added.



\begin{figure*}[t]

\begin{equation}
\textnormal{min}_{\Psi^{(k)} \in \mathbb{R}_+^{n \times m},\,\Theta^{(k)} \in \mathbb{R}_+^{m \times n}}\Bigg(\sum_{i=1}^n\sum_{j=1}^m \gamma_i \psi_{i,j}^{(k)}\sum_{p=1}^n \pi_{i,k} \textnormal{log}(\pi_{i,p}/\vartheta_{j,p}^{(k)})\,\Bigg|\!\begin{array}{c}0 \!\leq\! \vartheta_{i,q}^{(k)}, \psi_{i,q}^{(k)} \!\leq\! 1,\; \sum_{q=1}^m \vartheta_{i,q}^{(k)} \!=\! 1,\vspace{0.01cm}\\ \sum_{q=1}^m \psi_{i,q}^{(k)} \!=\! 1\,\, \forall i,j,k\end{array}\!\!\Bigg),\, \forall k
\end{equation}

\hrulefill

\begin{equation}
\textnormal{min}_{{\Psi^{(k)} \in \mathbb{R}_+^{n \times m}}}\left(\!\!\!\begin{array}{c}\mathbb{I}_{(k=0)}\sum_{i=1}^n\sum_{j=1}^m \gamma_i \psi_{i,j}^{(k)}\sum_{p=1}^n \pi_{i,p} \textnormal{log}(\pi_{i,p}/\vartheta_{j,p}^{(k)})\,+\vspace{0.05cm}\\ \mathbb{I}_{(k>0)}\sum_{q}\sum_{j=1}^m \omega_q \kappa_{q,j}^{(k)} \textnormal{log}(\kappa_{q,j}^{(k)}/\alpha_j^{(k)}) \end{array}\!\!\left|\!\begin{array}{c}\sum_{j=1}^m \alpha_j^{(k)} \sum_{i=1}^n \psi_{i,j}^{(k)}\textnormal{log}(\psi_{i,j}^{(k)}/\gamma_i) \leq r\vspace{0.05cm}\\ \kappa_{q,j}^{(k)} \!=\! \sum_{i=1}^n \gamma_i\, \eta_{q,i}\, \psi_{i,j}^{(k)}/\alpha_j^{(k)} \vspace{0.05cm}\\ 0 \!\leq\! \vartheta_{i,q}^{(k)}, \psi_{i,q}^{(k)} \!\leq\! 1,\; \sum_{q=1}^m \vartheta_{i,q}^{(k)} \!=\! 1,\, \sum_{q=1}^m \psi_{i,q}^{(k)} \!=\! 1\end{array}\!\!\!\right.\right)\!,\, \forall k
\end{equation}

\hrulefill\vspace{-0.5cm}

\end{figure*}

\begin{figure*}[b]
\vspace{-0.3cm}\hrulefill

\begin{equation}
\displaystyle\alpha_j^{(0)} \leftarrow \sum_{i=1}^n \gamma_i\,\psi^{(0)}_{i,j},\;\; \psi_{i,j}^{(1)} \leftarrow \!\Bigg(\alpha_j^{(0)} e^{-\beta \sum_{r=1}^n \pi_{i,r} \textnormal{log}(\pi_{i,r}/\vartheta_{j,r})}\!\Bigg)\!\Bigg/\!\Bigg(\sum_{p=1}^m \alpha_p^{(0)} e^{-\beta \sum_{r=1}^n \pi_{i,r} \textnormal{log}(\pi_{i,r}/\vartheta_{p,r})}\Bigg)\vspace{-0.05cm}
\end{equation}

\hrulefill

\begin{equation}
\displaystyle \alpha_j^{(k)} \leftarrow \sum_{i=1}^n \gamma_i\, \psi_{i,j}^{(k)},\;\; \kappa_{q,j}^{(k)} \leftarrow \sum_{i=1}^n \gamma_i\, \eta_{q,i}\, \psi_{i,j}^{(k)}/\alpha_j^{(k)},\;\; \psi_{i,j}^{(k+1)} \leftarrow \Bigg(\alpha_j^{(k)} e^{-\beta \sum_q \eta_{q,i}\textnormal{log}(\eta_{q,i}/\kappa_{q,j}^{(k)})}\!\Bigg)\!\Bigg/\!\Bigg(\sum_{p=1}^m \alpha_p^{(k)} e^{-\beta \sum_q \eta_{q,i}\textnormal{log}(\eta_{q,i}/\kappa_{q,p}^{(k)})}\!\Bigg)
\end{equation}
\end{figure*}

\section*{\textbf{2.$\;\;$METHODOLOGY}}\addtocounter{section}{1}

Our approach for aggregating Markov chains can be described as follows.  Given a stochastic matrix $\Pi \!\in\! \mathbb{R}^{n \times n}_+$ of transition probabilities between $n$ states, we seek to partition this matrix to produce a reduced-size stochastic matrix $\Phi \!\in\! \mathbb{R}^{m \times m}_+$ with $m$ states.  Since there are many possible $\Phi$'s that can be formed, we would like one with the least divergence to $\Pi$ for some measure.  Due to the different sizes of $\Pi$ and $\Phi$, though, directly assessing divergence is not possible.  To facilitate this comparison, we consider the construction of a joint-model stochastic matrix $\Theta \!\in\! \mathbb{R}^{m \times n}_+$ that encodes the dynamics of $\Phi$.

The definition of $\Theta$ relies on finding a partition matrix $\Psi \!\in\! \mathbb{R}^{m \times n}_+$ for determining which states in $\Pi$ should be combined to create an aggregated state in $\Phi$.  We show that an optimal $\Psi$ and $\Theta$, and hence $\Phi$, can be found by solving a modified value-of-information criterion.  We then characterize the errors associated with estimating this modified value-of-information for Markov chains with a finite number of states.  Removing the errors penalizes marginal improvements in the divergence associated with including more state groups $m$.  This non-linearly transforms the value of information such that it contains a global maximum for a single state group count.  We then specify an expression for this value, which coincides where the aggregation complexity is balanced against the preserved information.\vspace{0.15cm}



\noindent {{{\textbf{2.1$\;\;\;$Preliminaries}}}} \vspace{0.1cm}

For our approach, we consider a first-order, homogeneous Markov chain defined on a finite state space.  We assume that is nearly-completely decomposable.\vspace{0.15cm}  

\noindent{{{\textbf{Definition\! 2.1.}}}} The transition model of a first-order, homogeneous, nearly-completely-decomposable Markov chain is a weighted, directed graph $R_\pi$ given by the three-tuple $(V_\pi,E_\pi,\Pi)$ with:
\begin{itemize}
\item[] \-\hspace{0.5cm}(i) A set of $n$ vertices $V_\pi \!=\! v_\pi^1 \cup \ldots \cup v_\pi^n$ representing the states of the Markov chain.\vspace{-0.3cm}\\
\item[] \-\hspace{0.5cm}(ii) A set of $n \!\times\! n$ edge connections $E_\pi \subset V_\pi \!\times\! V_\pi$ between reachable states in the Markov chain.\vspace{-0.3cm}\\
\item[] \-\hspace{0.5cm}(iii) A stochastic transition matrix $\Pi \!\in\! \mathbb{R}_+^{n \times n}$.  Here, $[\Pi]_{i,j} \!=$\\ \noindent $\pi_{i,j}$ represents the non-negative transition probability between states $i$ and $j$.  We impose the constraint that the probability of experiencing a state transition is independent of time.  Moreover, for a block-diagonal matrix $\Pi^*$ with zeros along the diagonal, we have that $\Pi \!=\! \Pi^* \!+\! \varepsilon C$.  $\Pi^* \!\in\! \mathbb{R}_+^{n \times n}$ is a completely-decomposable stochastic matrix with $m$ indecomposable sub-matrix blocks $\Pi^*_i$ of order $n_i$.  The matrix $C \!\in\! \mathbb{R}_+^{n \times n}$ satisfies $\sum_{k=1}^{n_i}c_{p_i,k_i} \!=\! -\sum_{j \neq i}\sum_{q=1}^{n_j} c_{p_i,q_j}$ $\forall p_i$, for blocks $\Pi^*_i$ and $\Pi^*_j$.
\end{itemize}\vspace{0.15cm}

\noindent Throughout, we assume that all Markov chains are irreducible and aperiodic.  As a consequence, there is a unique invariant probability distribution $\gamma$ associated with the chain such that $\gamma^\top \Pi \!=\! \gamma^\top$.

We are interested in comparing pairs of nearly-completely-decomposable Markov chains.  A means to do this is by considering given rows of the stochastic transition matrix $\Pi$ with those of a reduced-order chain's stochastic matrix $\Phi$.  We will perform this comparison via the negative Kullback-Leibler divergence.  It coincides with the Donsker-Varadhan rate function appearing in the large-deviations theory of Markov chains \cite{DonskerMD-jour1975a,DonskerMD-jour1975b} and measures the dissimilarity between chains defined on the same discrete state space.



Since we are considering the problem of chain aggregation, the discrete state spaces will be different.  One chain $R_\pi$ will have $n$ states while another $R_\varphi$ will have $m$ states.  The dimensionalities of given rows in the corresponding transition matrices will hence not be equivalent, which precludes a direct comparison using the discrete Kullback-Leibler divergence.  To resolve this issue, we consider construction of a joint model $R_\vartheta$.  This joint model defines a joint state space composed of the states from $R_\pi$ and $R_\varphi$.  It, however, re-defines the edge set along with the weighting matrix.  This weighting matrix $\Theta$ has the same number of columns as $\Pi$ and the same dynamics as $\Phi$, which facilitates comparisons using conventional divergences.

The joint model relies on the specification of a partition matrix $\Psi$ for mapping states from the reduced-order model $R_\varphi$ to the original model $R_\pi$.  Here, we consider probabilistic partition functions so as to capture the inherent uncertainty in the state combination. \vspace{0.15cm}

\noindent{{{\textbf{Definition\! 2.2.}}}} Let $R_\pi \!=\! (V_\pi,E_\pi,\Pi)$ and $R_\varphi \!=\! (V_\varphi,E_\varphi,\Phi)$ be transition models of two Markov chains over $n$ and $m$ states, respectively.  A probabilistic partition function $\psi$ is a surjective mapping between two state index sets, $\mathbb{Z}_{1:n}$ and $\mathbb{Z}_{1:m}$, such that $\psi^{-1}(\mathbb{Z}_{1:m})$ is a partition of $\mathbb{Z}_{1:n}$, which has a given probabilistic chance of occurring.  That is, $\psi^{-1}(j) \!\subset\! \mathbb{Z}_{1:n} \times \mathbb{R}_+^n$ is not empty and where $\psi^{-1}(1) \cup \ldots \cup\, \psi^{-1}(m) \!=\!  \mathbb{Z}_{1:n}^m \times \mathbb{R}_+^{m \times n}$, with the real-valued responses being non-negative and summing to one.  

The probabilistic partition of a state index set induces a probabilistic partition matrix $[\Psi]_{i,j} \!=\! \psi_{i,j}$, where $\psi_{i,j} \!=\! \zeta$ if $i \!\in\! \psi^{-1}(j)$ occurs with probability $\zeta$.  The set of all probabilistic partition matrices is $\{\Psi \!\in\! \mathbb{R}_+^{n \times m}|[\Psi]_{i,j} \!=\! \psi_{i,j} \!\in\! [0,1],\; \sum_{j=1}^m \psi_{i,j} \!=\! 1\}$.\vspace{0.15cm}

\noindent{{{\textbf{Definition\! 2.3.}}}} Let $R_\pi \!=\! (V_\pi,E_\pi,\Pi)$ and $R_\varphi \!=\! (V_\varphi,E_\varphi,\Phi)$ be transition models of two Markov chains over $n$ and $m$ states, respectively, where $m \!<\! n$. $R_\vartheta \!=\! (V_\vartheta,E_\vartheta,\Theta)$ is a joint model, with $m \!+\! n$ states, that is defined by
\begin{itemize}
\item[] \-\hspace{0.5cm}(i) A vertex set $V_\vartheta \!=\! V_\pi \cup V_\varphi$, which is the union of all state vertices in $R_\pi$ and $R_\varphi$.
\item[] \-\hspace{0.5cm}(ii) An edge set $E_\vartheta \subset V_\varphi \!\times\! V_\pi$, which are one-to-many mappings from the states in the original transition model $R_\pi$ to the reduced-order transition model $R_\varphi$.  
\item[] \-\hspace{0.5cm}(iii) A weighting matrix $\Theta \!\in\! \mathbb{R}_+^{m \times n}$.  The partition function $\psi$ provides a relationship between the stochastic matrices $\Phi$ and $\Theta$ of $R_\varphi$ and $R_\vartheta$, respectively.  This is given by $\Phi \!=\! \Theta \Psi$, or, rather,\\ \noindent $\varphi_{i,j} \!=\! \sum_{k=1}^n \vartheta_{k,j}\psi_{k,i}$ $\forall i,j$, where $\Psi$ is a probabilistic partition matrix; here, $\psi_{i,j} \!=\! p(v_\varphi^j|v_\pi^i)$ and $\varphi_{i,j} \!=\! p(v_\varphi^j|v_\varphi^i)$.
\end{itemize}\vspace{0.15cm}

\noindent Our corresponding technical report should be consulted for further details about these choices and illustrations of the various concepts \cite{SledgeIJ-report2018a}.  This report also contains proofs for many of the ensuing claims.

\vspace{0.15cm}

\noindent {{{\textbf{2.2$\;\;\;$Aggregating\! Markov\! Chains}}}} \vspace{0.1cm}

For any given transition model $R_\pi$, we would like to find, by way of the joint model $R_\vartheta$, another transition model $R_\varphi$ with fewer states that resembles the dynamics encoded by $R_\pi$.  At the very least, a model $R_\vartheta$ should be sought with a weighting matrix $\Theta$ that has the least expected divergence with respect to the transition matrix $\Pi$ of $R_\pi$ for some partition.\vspace{0.15cm}

\noindent{{{\textbf{Definition\! 2.4.}}}} Let $R_\pi \!=\! (V_\pi,E_\pi,\Pi)$, $R_\varphi \!=\! (V_\varphi,E_\varphi,\Phi)$, and $R_\vartheta \!=\! (V_\vartheta,E_\vartheta,\Theta)$ be transition models of two Markov chains over $n$ and $m$ states and the joint model over $n \!+\! m$ states, respectively.  The least expected divergence between $\Pi$ and $\Theta$, and hence $\Pi$ and $\Phi$, is given in (1).  Here, $\gamma_i \!=\! p(v_\pi^i)$ and $\pi_{i,k} \!=\! p(v_\pi^i|v_\pi^k)$.\vspace{0.15cm}  

\noindent The least expected distortion possesses too few constraints to make discerning the number of state groups $m$ viable, though.  

\begin{figure*}[t]

\begin{equation}
\textnormal{min}_{{\Psi^{(k)} \in \mathbb{R}_+^{n \times m}}} \left(\!\!\!\!\left.\begin{array}{c}
\;\mathbb{I}_{(k=0)}\sum_{i=1}^n\sum_{j=1}^m \gamma_i \psi_{i,j}^{(k)}\sum_{p=1}^n \pi_{i,p} \textnormal{log}(\pi_{i,p}/\vartheta_{j,p}^{(k)}) + \mathbb{I}_{(k>0)}\sum_{q}\sum_{j=1}^m \omega_q \kappa_{q,j}^{(k)} \textnormal{log}(\kappa_{q,j}^{(k)}/\alpha_j^{(k)})\vspace{0.125cm}\\
+\, \mathbb{I}_{(k=0)}\sum_{g=2}^2 \frac{(-1)^g}{\textnormal{log}(2)(g^2 \!-\! g)} \sum_{j=1}^m \mathbb{E}[(\sum_{i=1}^n \overline{\gamma}_i\psi_{i,j}^{(k)} )^g]/(\alpha_j^{(k)})^{g-1} \vspace{0.125cm}\\
-\, \mathbb{I}_{(k>0)}\sum_{g=2}^\infty \frac{(-1)^g}{\textnormal{log}(2)(g^2 \!-\! g)}\sum_q \sum_{j=1}^m \frac{\omega_q^{g-1}\mathbb{E}[(\sum_{i=1}^n \overline{\eta}_{q,i} \rho_{i,j}^{(k)})^g] - (\rho_{q,j}^{(k)})^{g-1} \mathbb{E}[(\sum_{i=1}^n \gamma_y\overline{\eta}_{q,i})^g]}{(\rho_{q,j}^{(k)})^{g-1}\omega_q^{g-1}}\end{array}\!\!\right|\,\cdots\right)\!,\, \forall k
\end{equation}

\hrulefill\vspace{-0.5cm}

\end{figure*}

\begin{figure*}[b]
\vspace{-0.3cm}\hrulefill

\begin{equation}
\psi_{i,j}^{(1)} \leftarrow \!\Bigg(\sum_{y=1}^n \gamma_y\, \psi_{y,j}^{(0)}\, \textnormal{exp}\Bigg(\!\!-\! \beta \Bigg(\sum_{p=1}^n \pi_{i,p} \textnormal{log}\Bigg(\frac{\pi_{i,p}}{\vartheta_{j,p}^{(0)}}\Bigg)\!\Bigg) + \sum_{g=2}^\infty (-1)^g \sum_{p=1}^m \frac{\mathbb{E}[(\sum_{s=1}^n \overline{\gamma}_s\psi_{s,p}^{(0)})^g]}{g (\alpha_p^{(0)})^g} - \frac{\mathbb{E}[\,\overline{\gamma}_i (\sum_{s=1}^n \overline{\gamma}_s \psi_{s,p}^{(0)} )^{g-1}]}{(g \!-\! 1)\gamma_i (\alpha_p^{(0)})^{g-1} } \Bigg)\!\Bigg/\!\Bigg( \cdots \Bigg)\vspace{-0.05cm}
\end{equation}

\hrulefill

\begin{equation}
\!\!\!\!\!\begin{array}{c}

\displaystyle \alpha_j^{(k)} \leftarrow \sum_{i=1}^n \gamma_i\, \psi_{i,j}^{(k)},\;\; \kappa_{q,j}^{(k)} \leftarrow \sum_{i=1}^n \gamma_i\, \eta_{q,i}\, \psi_{i,j}^{(k)}/\alpha_j^{(k)},\;\;\overline{\kappa}_{q,j}^{(k)} \leftarrow \sum_{i=1}^n \overline{\eta}_{q,i}\, \tau_{i,j}^{(k)},\vspace{0.075cm}\\

\displaystyle\psi_{i,j}^{(k+1)} \leftarrow \!\Bigg(\alpha_j^{(k)} \textnormal{exp}\Bigg(\!\!-\!\beta \Bigg(\sum_q \eta_{q,i}\textnormal{log}\Bigg(\frac{\eta_{q,i}}{\kappa_{q,j}^{(k)}}\Bigg)\!\Bigg) \!-\! \beta\sum_{g=2}^\infty(-1)^g \Bigg(\sum_{q} \frac{\mathbb{E}[(\overline{\kappa}_{q,j}^{(k)})^g]}{\eta_{q,i}^{-1}\, g(\kappa_{q,j}^{(k)})^g} - \frac{\mathbb{E}[\overline{\eta}_{q,i}(\overline{\kappa}_{q,j}^{(k)})^{g-1}]}{(g \!-\! 1)(\kappa_{q,j}^{(k)})^g}\Bigg)\!\Bigg)\!\Bigg)\!\Bigg/\!\Bigg( \cdots \Bigg)
\end{array}\!\!\!\!\!
\end{equation}
\end{figure*}

To help discern $m$, we impose that the partitions should minimize the information loss associated with the state quantization process.  That is, the mutual dependence between states in the high-order and low-order chains should be maximized with respect to a supplied bound.  Simultaneously, the least expected divergence, for this bound, should be achieved while also seeking a maximal compression of the dynamics.  Realizing each of these competing objectives leads to a combined value-of-information and information-bottleneck aggregation process.\vspace{0.15cm}

\noindent{{{\textbf{Definition\! 2.5.}}}} Let $R_\pi \!=\! (V_\pi,E_\pi,\Pi)$, $R_\varphi \!=\! (V_\varphi,E_\varphi,\Phi)$, and $R_\vartheta \!=\! (V_\vartheta,E_\vartheta,\Theta)$ be transition models of two Markov chains over $n$ and $m$ states and the joint model over $n \!+\! m$ states, respectively.  An optimal reduced-order transition model $R_\varphi$ with respect to the original model $R_\pi$ can be found as follows:
\begin{itemize}
\item[] \-\hspace{0.5cm}(i) Optimal partitioning: Find a probabilistic partition matrix $\Psi$ that solves (2) for some  $r \!\in\! \mathbb{R}_+$; $r$ has an upper bound of $-\sum_{i=1}^n \gamma_i\textnormal{log}(\gamma_i)$.  As well, find the corresponding weighting matrix for $R_\vartheta$: $\Theta \!=\! U^\top \Pi$, where $[U]_{i,j} \!=\! \gamma_i \psi_{i,j}/\sum_{p=1}^n \gamma_p\psi_{p,j}$.
\item[] \-\hspace{0.5cm}(ii) Transition matrix construction: Obtain the transition matrix for $R_\varphi$ via: $\varphi_{i,j} \!=\! \sum_{k=1}^n \vartheta_{k,j}\psi_{k,i}$ using the optimal weights $\Theta$ and the probabilistic partition matrix $\Psi$ from step (i).
\end{itemize}

\noindent Here, $\alpha_j \!=\! p(v_\varphi^j)$, $\omega_q \!=\! p(q)$, $\eta_{q,i} \!=\! p(q|v_\pi^i)$, and $\kappa_{q,j} \!=\! p(q|v_\varphi^j)$ for some intermediate random variable $q$.\vspace{0.15cm}

A reason for considering this combined model is that simply trading off between the preserved information and the model complexity, as in the information bottleneck, does not encode notions of the underlying weighted-graph geometry.  This is because Shannon information is geometrically invariant: it yields an infinite set of degenerate solutions that do not minimize a specific divergence measure.  


Global solutions to (2) can be found via the initial condition in (3) and the expectation-maximization-like updates in (4).\vspace{0.15cm}
 
\noindent{{{\textbf{Proposition\! 2.1.}}}} Let $R_\pi \!=\! (V_\pi,E_\pi,\Pi)$ and $R_\varphi \!=\! (V_\varphi,E_\varphi,\Phi)$ be transition models of two Markov chains over $n$ and $m$ states.  The optimal partition matrix $\Psi$ that globally solves (2) can be found via the alternating update in (3) for iteration $k \!=\! 0$ and the alternating updates in (4) for iterations $k \!=\! 1,2,\ldots$  Here, $\beta \!\in\! \mathbb{R}_+$ is a Lagrange multiplier for handling the constraint bound $r \!\in\! \mathbb{R}_+$.\vspace{0.15cm}

\noindent The following proposition shows that the number of reduced-model state groups increases once $\beta$ reaches certain critical values.\vspace{0.15cm}

\noindent{{{\textbf{Proposition\! 2.2.}}}} Let $R_\pi \!=\! (V_\pi,E_\pi,\Pi)$ and $R_\varphi \!=\! (V_\varphi,E_\varphi,\Phi)$ be transition models of two Markov chains over $n$ and $m$ states.  For some $\beta_0$, suppose $\Theta_{\beta_0}$, the matrix $\Theta$ for that value of $\beta_0$, satisfies the following inequality $d^2/d\epsilon^2\, F(\Psi,K,\alpha;\Pi,\Theta_{\beta_0} \!+\! \epsilon Q,H,\gamma)|_{\epsilon = 0} > 0$, for the modified value-of-information Lagrangian from (2).  Here, $Q \!\in\! \mathbb{R}_+^{m \times n}$ is such that $\sum_{k=1}^m q_{k,1:n}^\top q_{k,1:n} \!=\! 1$ and $\sum_{j=1}^n q_{i,j} \!=\! 0$ $\,\forall i$.  A critical value $\beta_c$ satisfies $$\beta_c \!=\! \textnormal{min}_{\beta > \beta_0}\,(d^2/d\epsilon^2\, F(\Psi,K,\alpha;\Pi,\Theta_{\beta} \!+\! \epsilon Q,H,\gamma)|_{\epsilon = 0} \leq 0).$$  The number of rows in $\Theta$ and columns in $\Psi$ needs to be increased, by one, once $\beta \!>\! \beta_c$, since a phase change occurs.\vspace{0.125cm}

\noindent Sweeping over critical values of $\beta$ yields a hierarchy of probabilistic partitions $\Psi$ and hence reduced-order-model transition matrices $\Phi$ with different numbers of state groups.

This proposition does not, however, reveal a way to find the optimal number of state groups.  To do that, we will determine when the modified value-of-information begins to overfit to the transition-dynamics noise.  This leads to additional terms that we can subtract from the criterion to essentially regularize the aggregation and penalize for using too many or too few state groups than can be resolved in the transition dynamics. \vspace{0.15cm}

\noindent {{{\textbf{2.2.1$\;\;\;$Aggregated\! State\! Group\! Count}}}} \vspace{0.1cm}

We assume, for the modified value-of-information, that overfitting to noise is a byproduct of dealing with finite-state-space Markov chains and hence introducing estimation errors into the joint probabilities.  Characterizing the error effects and removing them yields a version of the value-of-information curve that is either convex or monotonically non-decreasing then monotonically non-increasing.\vspace{0.15cm} 

\noindent{{{\textbf{Definition\! 2.6.}}}}  Let $R_\pi \!=\! (V_\pi,E_\pi,\Pi)$, $R_\varphi \!=\! (V_\varphi,E_\varphi,\Phi)$, and $R_\vartheta \!=\! (V_\vartheta,E_\vartheta,\Theta)$ be transition models of two Markov chains over $n$ and $m$ states and the joint model over $n \!+\! m$ states, respectively.  An optimal reduced-order transition model $R_\varphi$, with respect to the original model $R_\pi$, can be found as follows:
\begin{itemize}
\item[] \-\hspace{0.5cm}(i) Optimal partitioning: Find a probabilistic partition matrix $\Psi$ that solves (5) for some $r \!\in\! \mathbb{R}_+$.  Equation (5) has the same constraint set as (2).  Find the corresponding weighting matrix for $R_\vartheta$: $\Theta \!=\! U^\top \Pi$, where $U$ is defined above.
\item[] \-\hspace{0.5cm}(ii) Transition matrix construction: Obtain the transition\ matrix for $R_\varphi$ via: $\varphi_{i,j} \!=\! \sum_{k=1}^n \vartheta_{k,j}\psi_{k,i}$ using the optimal weights $\Theta$ and the probabilistic partition matrix $\Psi$ from step (i).\vspace{0.025cm}
\end{itemize}

The systematic underestimation/overestimation error in (2) is $g$th-order minimized when solving (5).\vspace{0.15cm}

\noindent{{{\textbf{Proposition\! 2.3.}}}} Let $R_\pi \!=\! (V_\pi,E_\pi,\Pi)$ and $R_\varphi \!=\! (V_\varphi,E_\varphi,\Phi)$ be transition models of two Markov chains over $n$ and $m$ states.  The optimal partition matrix $\Psi$ for $R_\pi$ that globally solves (5) and facilitates the construction of $R_\varphi$ can be found via the updates in (6) for iteration $k \!=\! 0$ and the update in (7) for iterations $k \!=\! 1,2,\ldots$  For (6) and (7), the denominators are such that the partition-matrix entries are normalized to become probabilities.

Here, $\alpha_j \!=\! p(v_\varphi^j)$, $\gamma_i \!=\! p(v_\pi^i)$, $\psi_{i,j} \!=\! p(v_\varphi^j|v_\pi^i)$, $\eta_{q,i} \!=\! p(q|v_\pi^i)$, $\kappa_{q,j} \!=\! p(q|v_\varphi^j)$, $\tau_{i,j} \!=\! p(v_\pi^i|v_\varphi^j)$, and $\rho_{q,j} \!=\! p(q,v_\varphi^j)$.  The terms $\overline{\kappa}_{q,j}$ and $\overline{\eta}_{q,i}$ represent the estimation errors associated with $\kappa_{q,j}$ and $\eta_{q,i}$, while $\overline{\gamma}_i$ is the error associated with approximating $\gamma_i$.  We assume that the average error is zero: $\mathbb{E}[\overline{\eta}_{q,i}] \!=\! 0$ and $\mathbb{E}[\overline{\gamma}_{i}] \!=\! 0$.\vspace{0.15cm}

\begin{figure*}
   \hspace{-0.15cm}\begin{tabular}{c}
   \includegraphics[width=7.1in]{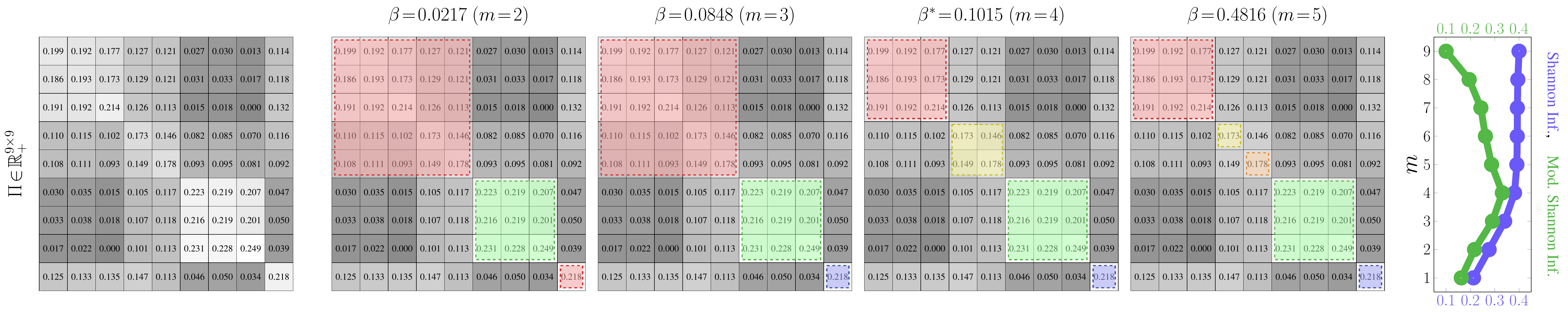}\vspace{-0.2cm}\\
   {\footnotesize (a)}\vspace{0.25cm}\\
   \includegraphics[width=7.1in]{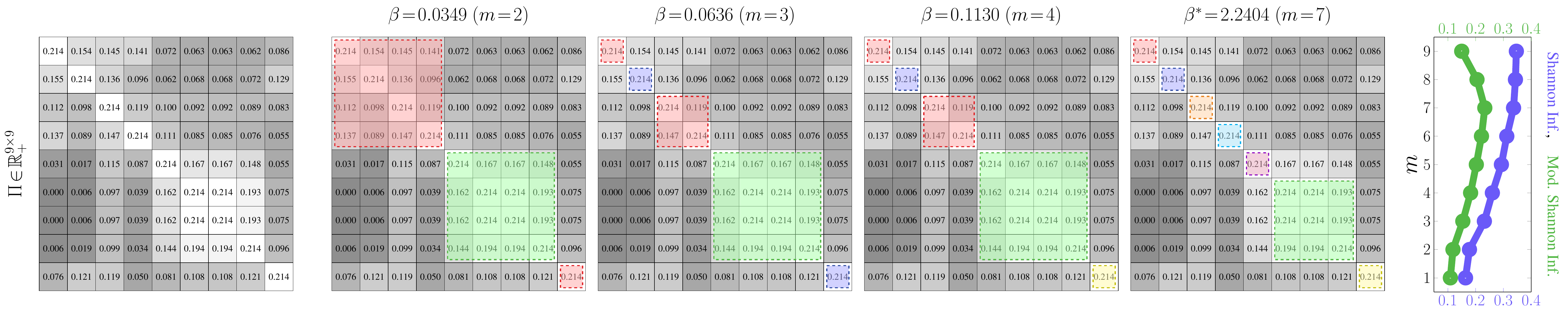}\vspace{-0.2cm}\\
   {\footnotesize (b)}
   \end{tabular}
\vspace{-0.025cm}
\caption[]{Modified value-of-information-based partitioning results for 9-state nearly-completely-decomposable Markov chains with: (a) four discernible state groups and (b) seven discernible state groups.  In both (a) and (b), we show the original stochastic matrix $\Pi$ with hardened versions of the partitions $\Psi$ overlaid for four critical values of $\beta$; as noted in the previous section, $m$ can be inferred from each value of $\beta$.  The unique colors in the partition plots correspond to which state group in $\Phi$ a state in $\Pi$ is most likely to be associated.  The right-most plots highlight the Shannon information in blue and the error-subtracted Shannon information in green.  After a certain number of clusters, the Shannon information plateaus, indicating that there is negligible benefit for including more state groups in $\Phi$.  The modified Shannon information begins to decrease when this occurs; the results align with the discernible number of groups for these chains.\vspace{-0.4cm}}
\end{figure*}

Due to the monotonicity properties of (5), we can construct upper and lower bounds for it and characterize their rate of change with respect to $\beta$.  This permits algebraically discerning where maxima of the equivalent dual problem (5) occur.\vspace{0.15cm}

\noindent{{{\textbf{Proposition\! 2.4.}}}}  Let $R_\pi \!=\! (V_\pi,E_\pi,\Pi)$, $R_\varphi \!=\! (V_\varphi,E_\varphi,\Phi)$, and $R_\vartheta \!=\! (V_\vartheta,E_\vartheta,\Theta)$ be transition models of two Markov chains over $n$ and $m$ states and the joint model over $n \!+\! m$ states, respectively.  The dual problem to (5) achieves a maximum for the Lagrange multiplier value $\beta^* \!=\! 2^{\sum_{i=1}^n\sum_{j=1}^m \gamma_i \psi_{i,j}\textnormal{log}(\psi_{i,j}/\alpha_j)}/2n$.\vspace{0.15cm}

\noindent As before, the partitioning process defined by (6) and (7) undergoes a series of phase changes whereby the number of state groups increase once $\beta$ exceeds some critical value.  The value of $\beta^*$ specified by the preceding proposition coincides with a partitioning where a certain number of state groups are defined.  It can be taken as the point in (5) where the complexity of the dynamics compression is balanced against the information contained about the original Markov chain in the reduced-order Markov chain.  For (2), $\beta^*$ often marks the beginning of the value-of-information's asymptotic region where there are diminishing returns for including more clusters.



\section*{\textbf{3.$\;\;$SIMULATIONS}}\addtocounter{section}{1}

In this section, we assess the empirical performance of the modified value-of-information criterion.  We gauge how well the predicted `optimal' free-parameter value aligns with the discernible number of state groups for nearly-completely-decomposable Markov chains.\vspace{0.15cm}

\noindent {{{\textbf{3.1$\;\;\;$Simulation\! Protocols}}}} \vspace{0.1cm}

We adopted the following simulation protocols for our aggregation framework.  We initialized the aggregation process with a partition matrix of all ones, $\Psi \!=\! [1]_{9 \times 1}$, signifying that each state belongs to a single group.  This is the global optimal solution of the aggregation problem and coincides with a parameter value $\beta$ of zero for the modified value-of-information.  We then found the subsequent critical values of $\beta$ and increased the column count of $\Psi$.  We determined which state group would be further split and modified both the new column and an existing column of $\Psi$ to randomly allocate the appropriate states.  This initialization process bootstraps the quantization for the new cluster and typically achieves convergence in only a few iterations.  It also permits the value of information to reliably track the global minimizer as $\beta$ increases.

For certain problems, a priori specifying a fixed amount of partition updates may not permit finding a steady-state solution.  We therefore run the alternating updates until no entries of the partition matrix change across two iterations.\vspace{0.15cm}

\noindent {{{\textbf{3.2$\;\;\;$Simulation\! Results\!}}}} \vspace{0.1cm}

We established the performance of the aggregation partitioning process through two examples.  The first, shown in figure 1(a), corresponds to a Markov chain with nine states and four state groups with strong intra-group interactions and weak inter-group interactions.  This is a relatively simple aggregation problem.  The second example, presented in figure 1(b), is of a nine-state Markov chain with a single dominant state group and six outlying states with near-equal transition probabilities.  This is a more challenging problem than the first, as the outlying states cannot be reliably combined without adversely impacting the mutual dependence.  

In figures 1(a) and 1(b), we provided partitions for four critical values of the free parameter $\beta$.  The `optimal' value of $\beta$, as predicted by our modified value of information, leads to four and seven state groups for the first and second examples, respectively.  Here, we considered a second-order approximation of the underestimation/overestimation error, i.e., $g \!=\! 2$.  Higher-order approximations did not modify the shape of the resulting Shannon-information curves much compared to those in figures 1(a) and 1(b); the resulting `optimal' number of state groups remained the same in these cases.  

For both examples, the associated partitions for the `optimal' values of $\beta$ align well with an inspection of the dynamics of the stochastic matrix: the partitions separate states that are more likely to transition to each other from those that are not.  Those partitions for `non-optimal' $\beta$'s either over- or under-quantize the chain states.  That is, for critical $\beta$'s before the `optimal' value, there is a moderate increase in the Shannon information, while the remaining $\beta$s only yield modest increases.  The `optimal' value of $\beta$ for both examples, in contrast, lies at the `knee' of this curve, which is where the divergence minimization is balanced against the competing objective of state-mutual-dependence maximization with respect to some bound.  It is also where the complexity of the aggregation result is in harmony with the information it contains about the original Markov chain.

To lend credence to these results, we considered forty reformulated graph-based partition validity measures \cite{SledgeIJ-jour2010a,SledgeIJ-jour2010b} across a hundred Monte Carlo simulations.  For the Markov chain in figure 1(a), thirty-three indices specified that there were four state groups while seven indices indicated that there were three state groups.  The results for the chain in figure 1(b) had more uncertainty, which was due to the high number of outlying states and the ability to combine them in multiple ways to reduce the models' divergence.  Ten indices suggested that there were two state groups, eight indices chose three state groups, while the remaining twenty-two favored either six or seven state groups almost equally.  

\renewcommand*{\bibfont}{\raggedright}
\renewcommand\bibsection{\section*{\textbf{REFERENCES}}}
{\setstretch{0.9}\selectfont \bibliography{sledgebib} \bibliographystyle{IEEEtran}}

\end{document}